\documentclass[twocolumn,prl,aps,amsmath,amssymb]{revtex4}
\usepackage{graphicx}
\usepackage{dcolumn}
\usepackage{bm}
\begin{document}
\title{
\hfill\vbox{\hbox{\small DPNU-03-22 \qquad\qquad}}\\
\hfill\vbox{\hbox{\small NORDITA-2003-58HE\qquad\qquad}}\\
\hfill\vbox{\hbox{\small SU-4252-787\qquad\qquad}}\\
Large $N_c$ and Chiral Dynamics }
 \author{Masayasu {\sc Harada}}
 \email{harada@eken.phys.nagoya-u.ac.jp}
 \affiliation{Department of Physics, Nagoya University, Nagoya 464-8602, Japan}
 \author{Francesco {\sc Sannino}}\email{francesco.sannino@nbi.dk}
 \affiliation{{\rm NORDITA} \& The Niels Bohr Institute, Blegdamsvej 17, DK-2100 Copenhagen \O,
Denmark }
\author{Joseph {\sc Schechter}}\email{schechte@physics.syr.edu}
 \affiliation{Department of Physics, Syracuse University,
Syracuse, NY 13244-1130, USA}
\date{September 2003}
\begin{abstract}
We study the dependence on the number of colors of the
 leading $\pi\pi$ scattering
amplitude in chiral dynamics. We demonstrate the existence of a
critical number of colors for and above which the low energy
$\pi\pi$ scattering amplitude computed from the simple sum of the
current algebra and vector meson terms
 is crossing symmetric and unitary at leading
order in a truncated and regularized $1/N_c$ expansion. The
critical number of colors turns out to be  $N_c=6$ and is
 insensitive to the explicit breaking
of chiral symmetry.
 Below this critical value, an additional state is
needed to enforce the unitarity bound; it is a broad one,
most likely of ``four quark" nature.
\end{abstract}
\maketitle
\section{Introduction}
\label{uno} Strong interactions are hard to tackle and different
tools or limits in the parameter space of a given asymptotically
free theory have been proposed. The large number of colors ($N_c$)
limit is a relevant example \cite{tH,Wi}. But how large should
 $N_c$ be for the
counting scheme to be not only qualitatively but also
quantitatively interesting for QCD ? In this note we try to answer this
question in terms of
 the nature of the low energy hadronic
spectrum.

The physical situation where the large $N_c$ approximation seems
closest to direct experimental check is perhaps low energy
$\pi\pi$ (or meson meson) scattering. Large $N_c$ predicts the low
energy amplitude to be a sum of tree diagrams -- contact terms and
resonance pole diagrams. With the two reasonable additions:
truncation of number of resonances and some kind of regularization
to give the resonances finite widths, this corresponds to the way
in which the analyses of experimentally derived partial wave
amplitudes are usually performed. We shall proceed in this manner
with the additional requirement that all pre-regularized tree
amplitudes be obtained from a
 chiral invariant (modulo usual symmetry breaking terms)
Lagrangian.

The most crucial partial wave for our considerations turns out to be
the scalar, isotopic spin zero channel.
Recently there has been renewed interest \cite{kyotoconf}-
\cite{BFMNS01} in the low energy behavior of this channel. Especially,
many physicists now believe in the existence of the light, broad
$I=J=0$ resonance, sigma in the 500-600 MeV region. The sigma is
also very likely the ``tip of an iceberg" which may consist of a
nonet of light four quark-type scalars \cite{Jaffe}
 mixing \cite{mixing} with another heavier
 $q \bar q$ - type scalar nonet as well as a
glueball. However there has been some controversy as to
 the underlying nature of
these low lying scalar objects. In the present context,
 it was argued \cite{Wi} that a four quark state in
the leading large $N_c$ limit would choose to appear as two
non-interacting $q{\bar q}$ states. The necessity of such a state
 to explain the scattering then suggests
a critical value of $N_c$, below which a simple analysis of $\pi
\pi$ scattering up to a given energy (taken as 1 GeV for
definiteness) and based on the large $N_c$ approximation is not
accurate.

To obtain this critical value, we make a
 reanalysis of meson-meson scattering
 as a function of the number of colors. The
$N_c$ dependences  can be easily
incorporated in the parameters (such as the pion decay constant or
the vector meson $\rho$ coupling to two pions) of the theory and
of the states whose large $N_c$ counting is known.
The criterion for an accurate description will be whether the
unitarity bounds are satisfied up to the energy of at least 1 GeV.

A variety of different approaches to meson-meson scattering have
been developed in the literature and all are able to provide a
reasonable fit to the data. An approach to describing $\pi\pi$,
$I=J=0$ scattering  based on a conventional non-linear chiral
Lagrangian of pseudoscalar and vector fields augmented by scalar
fields introduced in a chiral invariant manner was discussed in
\cite{{SS},HSS1} and will be adoped here. This will have an
advantage, for our present purpose, of not forcing the amplitudes
to be unitary. Thus unitarity can be considered as
a criterion for an
accurate theory.
  In the physical $N_c=3$ case, the experimental data from threshold
to a bit
beyond 1 GeV can be fit by computing the tree amplitude from this
Lagrangian and choosing parameters to yield
 an approximate unitarization. The following
ingredients are present: i) the ``current algebra" contact term,
ii) the vector meson exchange terms, iii) the unitarized
$\sigma$(560) exchange terms and iv) the unitarized $f_0$(980)
exchange terms. Although the $\rho$(770) vector meson certainly is
a crucial feature of low energy physics, a fit can be made
\cite{HSS2} in which the contribution ii) is absent. This results
in a somewhat lighter and broader sigma meson, in agreement with
other approaches which neglect the effect of the $\rho$ meson. We
note that the $\sigma(560)$ contribution is needed to keep the
$I=J=0$ scattering amplitude within the unitarity bounds
\cite{SS}. In \cite{HSS1} it was noted that the $\sigma(560)$
contribution is also needed to provide the correct background
which flips the sign of the $f_0(980)$ resonance (the
Ramsauer-Townsend effect).

For the purpose of checking the above strong interaction calculation,
the meson-meson scattering was also calculated in a general
version of the {\it linear} SU(3) sigma model, which contains both
$\sigma$(560) and $f_0(980)$ candidates \cite{BFMNS01}.
 The procedure adopted was to calculate the
tree amplitude and then to unitarize, without introducing any new
parameters, by identifying the tree amplitude as the K-matrix
amplitude. This also gave a reasonable fit to the data, including
the characteristic Ramsauer-Townsend effect of the $f_0(980)$
resonance contribution. At a deeper and more realistic level of
description in the linear sigma model framework, one expects two
different chiral multiplets - $q{\bar q}$ as well as $qq{\bar
q}{\bar q}$ - to mix with each other. A start on this model seems
encouraging \cite{sectionV}.

Although we shall adopt the non-linear realization approach
\cite{SS,{HSS1}}, our results are expected to be
robust. We will
also focus, for simplicity, on the SU(2) flavor limit of QCD.
 In section II we study the real parts of
the $\pi\pi$ scattering amplitudes for the $I=J=0$ and $I=J=1$
channels as  functions of the number of colors. The contributions
of the current algebra and the vector meson $\rho$ exchange pieces
are both taken into account. We show that chiral symmetry,
crossing symmetry and the unitarity bounds are all satisfied in
the leading order in $N_c$ for $N_c\geq 6$. However the unitarity
bound will not hold for lower values of $N_c$. Section III is
dedicated to the low energy scalar (the $\sigma$) which is needed
to unitarize the $R^0_0$ amplitude for $N_c\leq 6$.

\section{Approaching the large $N_c$ limit in $\pi\pi$ Scattering }
\label{due}

We first analyze the $I=J=0$ pion-pion scattering amplitude as a
function of the number of colors. In terms of the conventional
amplitude, $A(s,t,u)$ the $I=0$ amplitude is
$3A(s,t,u)+A(t,s,u)+A(u,t,s)$. One gets the $J=0$ state by
projecting out the correct partial wave. We focus, following
\cite{SS,{HSS1}}, on the real part of the scattering amplitude
since the imaginary part can be determined, using the unitarity
relation, from the real part. Of course, unitarity requires that
the real part of the elastic scattering amplitude itself must
satisfy the constraint $|R^I_J|\leq 1/2$. This condition
is crucial to obtain the critical number of colors in the
following analysis.

\begin{figure}[htpb]
\centering
{\includegraphics[width=6cm,clip=true]{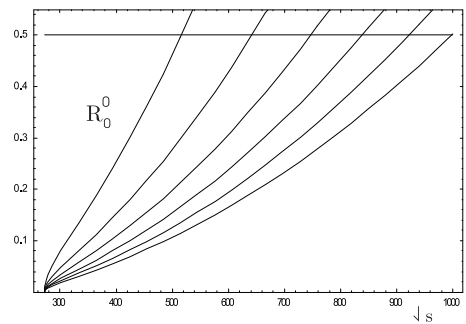}}
\caption[] {Real part of the $I=J=0$ partial wave amplitude due to
the current algebra term plotted for the following increasing
values of $N_c$ (from left to right), $3,5,7,9,11,13$.} \label{ca}
\end{figure}

The current algebra contribution to the
conventional amplitude is
\begin{eqnarray}
A_{ca}(s,t,u)=2\frac{s-m^2_{\pi}}{F^2_{\pi}}\ ,
\label{eq:ca}
\end{eqnarray}
where the pion decay constant, $F_{\pi}$ depends on $N_c$ as
$F_{\pi}(N_c)=131\sqrt{N_c}/\sqrt{3}$ so that $F_{\pi}(3)=131$~MeV.
Furthermore $m_{\pi}=137$~MeV is independent of $N_c$.
 In Fig.~\ref{ca} we plot the current algebra contribution to
 the real part of the $I=J=0$
partial wave amplitude, $R^0_0$
 for increasing values of the
number of colors $N_c$. We observe that the unitarity bound is
violated for a value of $s^{\ast}_{ca}$ which increases linearly
with $N_c$.
Thus the net effect of increasing $N_c$ is to delay the onset of
the  unitarity
bound violation. For $N_c=13$  the unitarity bound is
satisfied  up to about
1~GeV. Thus the current algebra contribution
 alone indicates that $N_c$ should be of
the order 13 to lead to a unitary chiral theory.

We now demonstrate that this result is strongly
modified by the presence of the well established $q\bar{q}$
companion of the pion -- the $\rho$ vector meson. The amplitude
we consider now is obtained by adding to the current algebra
contribution the following vector meson $\rho(770)$ contribution:
\begin{eqnarray}A_{\rho}(s,t,u)&=&
\frac{g_{\rho\pi\pi}^2}{2m^2_{\rho}}(4m^2_{\pi}-3s)\nonumber
\\&-&\frac{g_{\rho\pi\pi}^2}{2}\left[\frac{u-s}{(m^2_{\rho}-t)
-im_{\rho}\Gamma_{\rho}\theta(t-4m^2_{\pi})}
\right.\nonumber \\&+& \left.
\frac{t-s}{(m^2_{\rho}-u)-im_{\rho}\Gamma_{\rho}\theta(u-4m^2_{\pi})}\right]
\ ,\label{arho}
\end{eqnarray}
where $g_{\rho\pi\pi}(N_c)= 8.56
\sqrt{3}/\sqrt{N_c}$ is the $\rho \pi \pi$ coupling constant.
Also,
$m_{\rho}=771$~MeV is independent of $N_c$ and \begin{eqnarray}
\Gamma_{\rho}(N_c)=\frac{g^2_{\rho
\pi\pi}\left(N_c\right)}{12\pi\,m_{\rho}^2}
\left(\frac{m^2_{\rho}}{4} - m_{\pi}^2 \right)^{\frac{3}{2}} \ .
\end{eqnarray}
It should be noted that the first term in Eq.~(\ref{arho}),
which implies the existence of
an additional non-resonant contact interaction other than the current
algebra contribution, is required by the chiral symmetry when we
include the $\rho$ vector meson contribution in a chiral invariant
manner.
Although the vector meson contribution to the $I=J=0$
partial wave amplitude is due only to the
 crossed channel, as noted in Ref.~\cite{SS},
adding the $\rho$(770)  for the 3 color case greatly
decreases the amount of unitarity violation. However
for three colors this contribution is
not sufficient to
keep the scattering amplitude completely within the unitarity
bounds and a broad state (the sigma) is needed~\cite{SS}.

\begin{figure}[htbp]
\centering \leavevmode {\includegraphics[width=
6cm,clip=true]{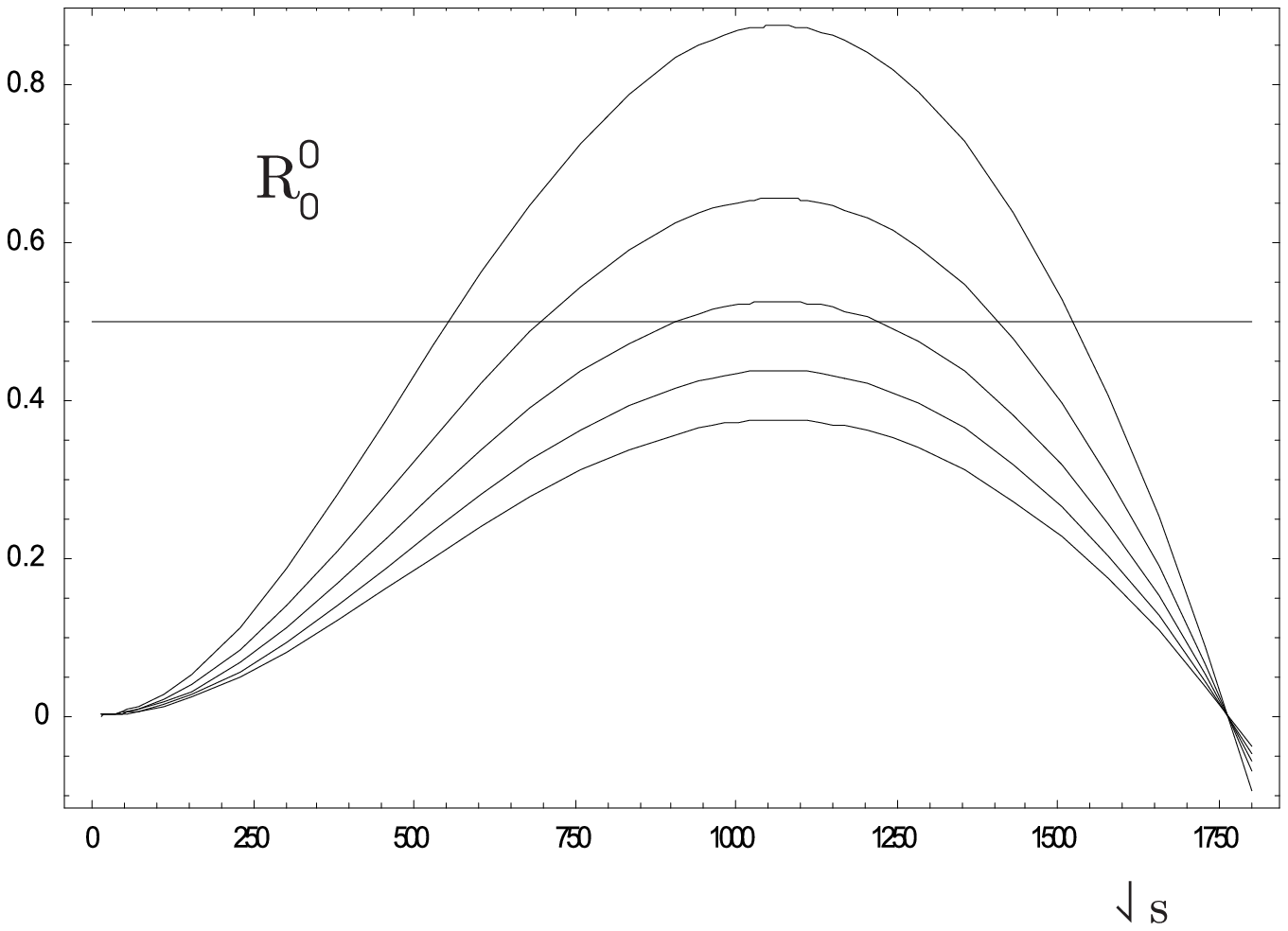}} \leavevmode {
\includegraphics[width=6cm,clip=true]{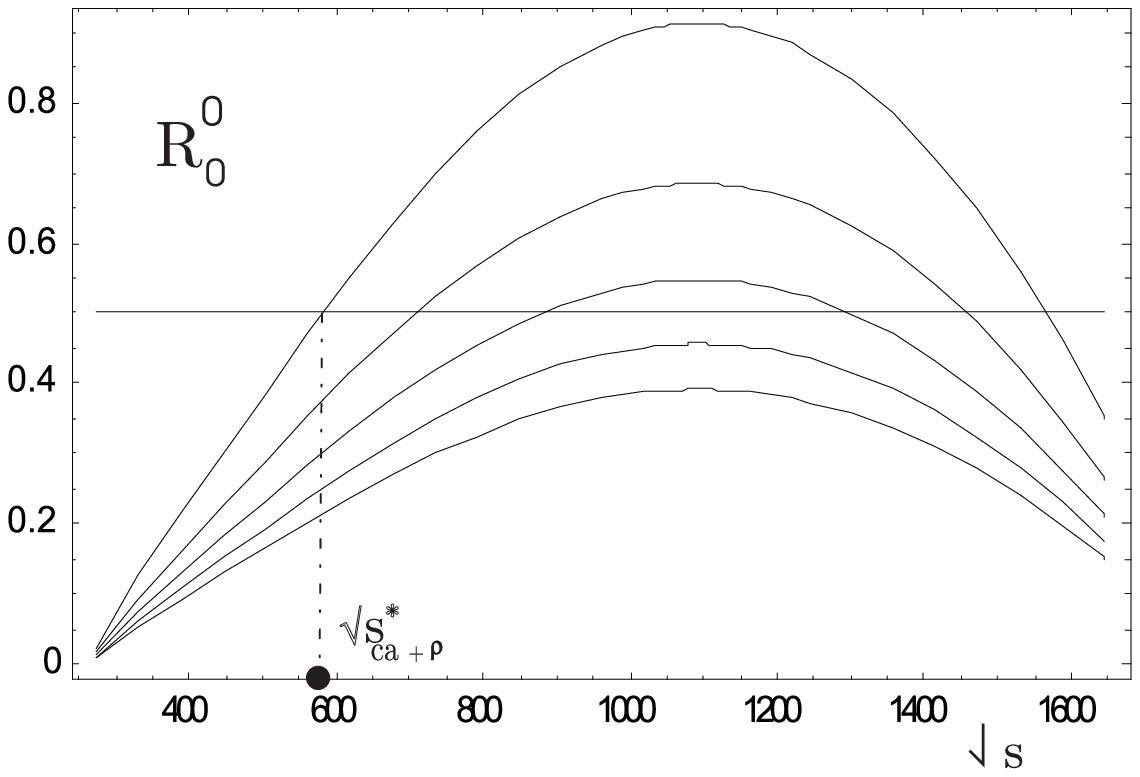}} \caption[]
{Upper Panel: Real part of the $I=J=0$ partial wave amplitude due
to the current algebra $+\rho$ terms, plotted for the following
increasing values of $N_c$ (from up to down),
$3,4,5,6,7$. The curve with largest
violation of the unitarity bound corresponds to $N_c=3$ while the
ones within the unitarity bound are for $N_c=6,7$. Here we have
set $m_{\pi}=0$. Lower Panel:
Same plot for the physical value of
the pion mass.} \label{carho}
\end{figure}

Here we follow a different route and investigate the unitarity bound
restoration --in the absence of the sigma-- as a function of $N_c$.
We compute analytically the relevant partial wave projections.
{}In
Fig.~\ref{carho} we plot the real part of the $I=J=0$
amplitude (i.e. $R^0_0$) due to current algebra plus the $\rho$
contribution for increasing values of $N_c$. Since in this channel
the vector meson is never on shell we suppress the contribution of
the width in the vector meson propagator in Eq.~(\ref{arho}).
We
observe that the unitarity bound (i.e., $\vert R_0^0 \vert \le 1/2$)
 is satisfied for $N_c\geq 6$ till well beyond
the 1~GeV region. However unitarity is still a problem for $3,4$
and $5$ colors. {}At energy scales larger than the one associated
with the vector meson clearly other resonances are needed \cite{SS}
but we shall not be concerned about those here. It is also
interesting to note that these considerations are essentially
unchanged when the pion mass (i.e. explicit chiral symmetry
 breaking in the Lagrangian) is set to zero.
We also note that the curves in Fig.~\ref{carho} are
multiples of each other,
so all of the extrema coincide as well as the points where the
curves cross zero. This is so since the
$N_c$ dependent scattering amplitude which
is the sum of the current algebra and vector meson contributions
obeys, by construction:
\begin{eqnarray}
A(s,t,u)=\frac{3}{N_c}\widetilde{A}(s,t,u) \ .
\label{Nc scaling}
\end{eqnarray}
where $\widetilde{A}(s,t,u)$ is defined replacing $F_{\pi}$ and
$g_{\rho\pi\pi}$ with the $N_c$ independent quantities
$\widetilde{F}_{\pi}= F_{\pi}\,\sqrt{3}/\sqrt{N_c}$ and
$\widetilde{g}_{\rho\pi\pi}=g_{\rho\pi\pi}\sqrt{N_c}/\sqrt{3}$~
~\footnote{
 Strictly speaking, the $N_c$ scaling property in
 Eq.~(\ref{Nc scaling}) is true for $t<0$ and $u<0$ where the
 width in the vector meson propagator in Eq.~(\ref{arho}) vanishes.
}.

\begin{figure}[thpb]
\centering
{\includegraphics[width=7cm,clip=true]{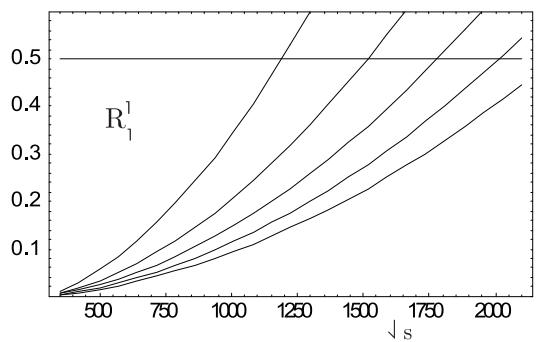}} \caption[] {
Real part of the $I=J=1$ partial wave amplitude due to current
algebra plotted for the following increasing values of $N_c$
(from left to right),
$3,5,7,9,11$. The curve which first violates the unitarity bound
corresponds to $N_c=3$.} \label{ca11}
\end{figure}

\begin{figure}[thpb]
\centering
{\includegraphics[width=7cm,clip=true]{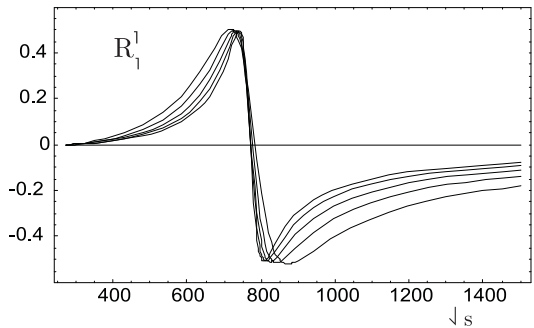}} \caption[]
{Real part of the $I=J=1$ partial wave amplitude due to the
current algebra $+\rho$ terms, plotted for the following increasing
values of $N_c$, $3,4,5,6,7$. $N_c=3$ is the curve with the
broadest $\rho$ meson.} \label{carho11}
\end{figure}

One might be worried that the critical value, $N_c=6$ determined by
studying the $I=J=0$ channel would not hold for other
channels. This is not so since $I=J=0$ is actually the worst
channel with respect to unitarity bound violation at the low
energies of present interest. Indeed channels with
 larger $J$ have angular
momentum
suppression factors which delay the onset of such a violation.
Let us consider as a relevant example the $I=J=1$ channel. Figure
\ref{ca11} shows that, even including just the current algebra term,
the unitarity bound is not an issue for 3
colors till energies well above
the $\rho$ mass.
The combined contributions of the vector meson $\rho$ term
and the current algebra term in this channel as a function of the number
of colors is shown in Fig.~\ref{carho11}. For this channel the
non zero $\rho$ width has been of course taken into acount. The already
narrow
$\rho$ becomes more and more narrow as we increase $N_c$ while the
$N_c$ effects are not very relevant concerning unitarity bounds.
Near the vector resonance we observe practically a Breit-Wigner
consistent with the well known cancellation between the current algebra
contribution and part of the non resonant vector meson
contribution in chiral models.

It is amusing that the vector meson plays a major role in
helping satisfy the unitarity bound for large $N_c$ not
in the direct channel but in the crossed one. It is also
interesting that the critical number of colors
emerges from the relation between $g_{\rho \pi \pi}$,
$F_{\pi}$ and $m_{\rho}$
while explicit chiral symmetry breaking has been seen
to play a negligible role. Notice that in the most
crucial $I=J=0$ channel the vector meson width can
be deleted to good accuracy and plays no role
in satisfying the unitarity bound. Thus apart from the
truncation of higher mass resonant states the calculation of this
partial
wave is exactly that of the leading large $N_c$ approximation.
We also note that the $I=2, J=0$ partial wave amplitude
computed in this way obeys the unitarity bound to well above
1 GeV \cite{Iequals2}.

{}We have thus demonstrated that for $N_c\geq 6$ and two flavor
QCD the low energy theory is crossing symmetric and unitary up to
well beyond the one GeV region considering just the pion and the
$\rho$ vector meson. This  supports the usual expectation that
when $N_c$ is large {\it enough} the meson-meson scattering theory
is well represented by including just the relevant $q\bar{q}$
resonances at the tree level.

\section{Including the sigma}
\label{tre}

In order to explain low energy $\pi\pi$ scattering for the physical
value
$N_c=3$ (and for the cases $N_c=4,5$) using the attractive notion
of tree diagram dominance involving near by resonances, it thus
seems necessary to include a scalar singlet resonance like the
light sigma which is likely to be of ``four quark" type and hence
not to contribute in the leading large $N_c$ approximation. As
mentioned before, many models \cite{kyotoconf}-\cite{BFMNS01} have
recently been proposed which successfully employ the sigma in one
way or another. Some discussions of $N_c$ dependence have also
been very recently given \cite{Pelaez:2003rv}.

\begin{figure}[thpb] \centering
{\includegraphics[width=6cm,clip=true]{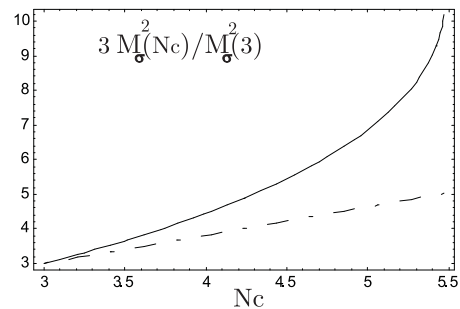}} \caption[]
{The value $M^2(N_c)$ for which unitarity bounds are first
violated as function of $N_c$ and normalized to $M^2(3)/3$. The
dashed line corresponds to the pure current algebra contribution
while the solid line to current algebra $+\rho$ contribution.}
\label{Masse}
\end{figure}

    Using the approach of Refs.~\cite{SS,HSS1} it is easy to
make a quick estimate of the mass
of the needed sigma particle.
A unitarization procedure is needed when the
current algebra plus the $\rho$ amplitude first starts violating
the unitarity bound. Denote this point as $s^{\ast}$.
 The $\sigma$ pole structure is such that the
real part of its amplitude is positive for $s<M^2_{\sigma}$ and
negative for $s>M^2_{\sigma}$. Identifying the squared sigma mass roughly
with $s^{\ast}$ will then give a negative contribution where
the real part of the amplitude exceeds $+1/2$. In the
 case when only the current algebra term is included we get
\begin{eqnarray}
M^2_{\sigma} \approx s^{\ast}_{ca}=4\pi\,F^2_{\pi} \ .
\end{eqnarray}
This shows that the squared mass of the sigma meson
needed to restore unitarity for $N_c=3,4,5$ increases
 roughly linearly with $N_c$. This estimate gets modified a bit
when we include the vector meson (see Fig.~{\ref{Masse}}),
yielding
 $M^2_{\sigma} \approx
s^{\ast}_{ca+\rho}$, where $s^{\ast}_{ca+\rho}$ is to be obtained
from Fig.~\ref{carho}.
 In the case of
 three colors we then get $M_{\sigma}
\approx 580$~MeV which is close to the best fit to the $\pi\pi$
scattering data determined in Refs.~\cite{SS,HSS1}.

    Finally, we briefly discuss the possible four quark nature of the
sigma. Most naively, the  needed particle is rather light while a
usual constituent quark $q{\bar q}$ bound state should be a
somewhat heavy p-wave state. Of course we have been working in an
effective Lagrangian framework which does not directly tell us
anything about the inner structure of the sigma meson. Indirectly
one can examine the three flavor generalization and observe that
the lightest nine scalars found in the present manner have the
right flavor  quantum numbers to make up a usual nonet. However
the characteristics of this nonet, such as usual vs. inverted mass
ordering and the value of the mixing angle between the two
isoscalars, turn out~\cite{nonet} to favor a four quark
interpretation. Still the four quarks may be of $qq{\bar q}{\bar
q}$ type \cite{Jaffe}, meson-meson molecule type \cite{molecule}
or any linear combination. For our present purpose it is enough to
know that they are not of $q{\bar q}$ type.
\section{Conclusions}
We studied the dependence on number of colors
of $\pi\pi$ scattering. We showed the existence of a critical
number of colors for and above which, chiral theory is unitary at
leading order in the $1/N_c$ expansion to beyond the one GeV
range. The critical number of colors is $N_c=6$ and is insensitive
to the
 explicit breaking of chiral
symmetry. For $N_c\geq6$
the $q\bar{q}$ resonances are sufficient to keep the low energy theory
unitary. Below this critical value the $\sigma$ meson is needed to
keep
the amplitude within the unitarity bounds.
 This is a light, broad object which is unlikely to be of
$q{\bar q}$ nature.

 \acknowledgments
It is a pleasure to thank A. Abdel Rehim, D. Black, A. H.
Fariborz, A. M\'{o}csy, S. Nasri and K. Tuominen for discussions
and for careful reading of the manuscript. The work of F.S. is
supported by the Marie--Curie fellowship under contract
MCFI-2001-00181. The work of J.S. is supported in part by the U.
S. DOE under Contract no. DE-FG-02-85ER 40231.


\begin{thebibliography}{9}

\bibitem{tH} G.'t Hooft, Nucl. Phy. {\bf B72}, 461 (1974).

\bibitem{Wi} E. Witten, Nucl. Phy. {\bf B160}, 57 (1979).

\bibitem{kyotoconf}
See the dedicated conference proceedings, S. Ishida et al
``Possible existence of the sigma meson and its implication to
hadron physics", KEK Proceedings 2000-4, Soryyushiron Kenkyu 102,
No. 5, 2001. Additional points of view are expressed in the
proceedings, D. Amelin and A.M. Zaitsev ``Hadron Spectroscopy'',
Ninth International Conference on Hadron Spectroscopy, Protvino,
Russia(2001).

\bibitem{vanBev}
E.~Van Beveren, T.~A.~Rijken, K.~Metzger, C.~Dullemond, G.~Rupp and
J.~E.~Ribeiro,
Z.\ Phys.\ C {\bf 30}, 615 (1986);
E.~van Beveren and G.~Rupp,
Eur.\ Phys.\ J.\ C {\bf 10}, 469 (1999).
See also
J.~J.~De Swart, P.~M.~Maessen and T.~A.~Rijken,
Talk given at U.S. / Japan Seminar: The Hyperon - Nucleon Interaction,
Maui, HI, 25-28 Oct 1993
[arXiv:nucl-th/9405008].

\bibitem{MP}
D.~Morgan and M.~R.~Pennington,
Phys.\ Rev.\ D {\bf 48}, 1185 (1993).

\bibitem{BMPV}
A.~A.~Bolokhov, A.~N.~Manashov, V.~V.~Vereshagin and V.~V.~Polyakov,
Phys.\ Rev.\ D {\bf 48}, 3090 (1993).
See also
V.~A.~Andrianov and A.~N.~Manashov,
Mod.\ Phys.\ Lett.\ A {\bf 8}, 2199 (1993).
Extension of this string-like approach to the $\pi K$
case has been made in
V.~V.~Vereshagin,
Phys.\ Rev.\ D {\bf 55}, 5349 (1997)
and  in
A.~V.~Vereshagin and V.~V.~Vereshagin,
Phys.\ Rev.\ D {\bf 59}, 016002 (1999).

\bibitem{AS94}
N.~N.~Achasov and G.~N.~Shestakov,
Phys.\ Rev.\ D {\bf 49}, 5779 (1994).

\bibitem{Kam94}
R.~Kam\'inski, L.~Le\'sniak and J.~P.~Maillet,
Phys.\ Rev.\ D {\bf 50}, 3145 (1994).

\bibitem{SS}
F.~Sannino and J.~Schechter,
Phys.\ Rev.\ D {\bf 52}, 96 (1995).

\bibitem{T}
N.~A.~Tornqvist,
Z.\ Phys.\ C {\bf 68}, 647 (1995)
and references therein.
In addition see
N.~A.~T\"ornqvist and M.~Roos,
Phys.\ Rev.\ Lett.\  {\bf 76}, 1575 (1996);
N.~A.~Tornqvist,
Talk given at 7th International Conference on Hadron Spectroscopy
(Hadron 97), Upton, NY, 25-30 Aug 1997 and at EuroDaphne Meeting,
Barcelona, Spain, 6-9 Nov 1997
[arXiv:hep-ph/9711483];
Phys.\ Lett.\ B {\bf 426}, 105 (1998).

\bibitem{DS}
R.~Delbourgo and M.~D.~Scadron,
Mod.\ Phys.\ Lett.\ A {\bf 10}, 251 (1995);
See also
D.~Atkinson, M.~Harada and A.~I.~Sanda,
Phys.\ Rev.\ D {\bf 46}, 3884 (1992).

\bibitem{JPHS}
G.~Janssen, B.~C.~Pearce, K.~Holinde and J.~Speth,
Phys.\ Rev.\ D {\bf 52}, 2690 (1995).

\bibitem{Sv}
M.~Svec,
Phys.\ Rev.\ D {\bf 53}, 2343 (1996).


\bibitem{Ishida}
S.~Ishida, M.~Ishida, H.~Takahashi, T.~Ishida, K.~Takamatsu and
T.~Tsuru,
Prog.\ Theor.\ Phys.\  {\bf 95}, 745 (1996);
S.~Ishida, M.~Ishida, T.~Ishida, K.~Takamatsu and T.~Tsuru,
Prog.\ Theor.\ Phys.\  {\bf 98}, 621 (1997).
See also
M.~Ishida and S.~Ishida,
Talk given at 7th International Conference on Hadron Spectroscopy
(Hadron 97), Upton, NY, 25-30 Aug 1997
[arXiv:hep-ph/9712231].


\bibitem{HSS1}
M.~Harada, F.~Sannino and J.~Schechter,
Phys.\ Rev.\ D {\bf 54}, 1991 (1996).

\bibitem{HSS2}
M.~Harada, F.~Sannino and J.~Schechter,
Phys.\ Rev.\ Lett.\  {\bf 78}, 1603 (1997).

\bibitem{BFSS1}
D.~Black, A.~H.~Fariborz, F.~Sannino and J.~Schechter,
Phys.\ Rev.\ D {\bf 58}, 054012 (1998).

\bibitem{BFSS2}
D.~Black, A.~H.~Fariborz, F.~Sannino and J.~Schechter,
Phys.\ Rev.\ D {\bf 59}, 074026 (1999).

\bibitem{OOP}
J.~A.~Oller, E.~Oset and J.~R.~Pelaez,
Phys.\ Rev.\ Lett.\  {\bf 80}, 3452 (1998).
See also
K.~Igi and K.~i.~Hikasa,
Phys.\ Rev.\ D {\bf 59}, 034005 (1999).

\bibitem{AnSa}
A.~V.~Anisovich and A.~V.~Sarantsev,
Phys.\ Lett.\ B {\bf 413}, 137 (1997).

\bibitem{EFSS}
V.~Elias, A.~H.~Fariborz, F.~Shi and T.~G.~Steele,
Nucl.\ Phys.\ A {\bf 633}, 279 (1998).

\bibitem{Dm}
V.~Dmitrasinovic,
Phys.\ Rev.\ C {\bf 53}, 1383 (1996).

\bibitem{MO}
P.~Minkowski and W.~Ochs,
Eur.\ Phys.\ J.\ C {\bf 9}, 283 (1999).

\bibitem{GN}
S.~Godfrey and J.~Napolitano,
Rev.\ Mod.\ Phys.\  {\bf 71}, 1411 (1999).

\bibitem{BG}
L.~Burakovsky and T.~Goldman,
Phys.\ Rev.\ D {\bf 57}, 2879 (1998).

\bibitem{FS1}
A.~H.~Fariborz and J.~Schechter,
Phys.\ Rev.\ D {\bf 60}, 034002 (1999).

\bibitem{BFS2}
D.~Black, A.~H.~Fariborz and J.~Schechter,
Phys.\ Rev.\ D {\bf 61}, 074030 (2000).
See also
V.~Bernard, N.~Kaiser and U.~G.~Meissner,
Phys.\ Rev.\ D {\bf 44}, 3698 (1991).

\bibitem{BFS3}
D.~Black, A.~H.~Fariborz and J.~Schechter,
Phys.\ Rev.\ D {\bf 61}, 074001 (2000).

\bibitem{Shakin}
L.~S.~Celenza, S.~f.~Gao, B.~Huang, H.~Wang and C.~M.~Shakin,
Phys.\ Rev.\ C {\bf 61}, 035201 (2000).

\bibitem{BFMNS01}
D.~Black, A.~H.~Fariborz, S.~Moussa, S.~Nasri and J.~Schechter,
Phys.\ Rev.\ D {\bf 64}, 014031 (2001).

\bibitem{Jaffe}
R.~L.~Jaffe,
Phys.\ Rev.\ D {\bf 15}, 267 (1977);
Phys.\ Rev.\ D {\bf 15}, 281 (1977).

\bibitem{mixing}
In addition to \cite{BFS3} and \cite{BFMNS01}
above see
T.~Teshima, I.~Kitamura and N.~Morisita,
J.\ Phys.\ G {\bf 28}, 1391 (2002);
F. Close and N. Tornqvist, {\it ibid.} {\bf
28}, R249 (2002) and
A.~H.~Fariborz,
arXiv:hep-ph/0302133.

\bibitem{sectionV}
See section V of \cite{BFMNS01} above.

\bibitem{Iequals2}
See \cite{SS} above and also
M.~S.~Chanowitz and W.~Kilgore,
Phys.\ Lett.\ B {\bf 322}, 147 (1994).

\bibitem{Pelaez:2003rv}
J.~R.~Pelaez,
arXiv:hep-ph/0306063;
M. Uehara, arXiv:hep-ph/0308241.

\bibitem{nonet}  Black et al \cite{BFSS2} above;
V.~Cirigliano, G.~Ecker, H.~Neufeld and A.~Pich,
JHEP {\bf 0306}, 012 (2003);
J.~A.~Oller,
arXiv:hep-ph/0306031.

\bibitem{molecule}
J.~D.~Weinstein and N.~Isgur,
Phys.\ Rev.\ Lett.\  {\bf 48}, 659 (1982).

\end{thebibliography}
\end{document}